\documentclass[twocolumn,tighten]{aastex631}
\usepackage{graphicx}
\usepackage{color}

\begin{document}

\shortauthors{Luhman et al.}
\shorttitle{Brown Dwarfs in the ONC}

\title{JWST/NIRSpec Observations of Brown Dwarfs in the Orion Nebula
Cluster\footnote{Based on observations made with the NASA/ESA/CSA James 
Webb Space Telescope.}}

\author{K. L. Luhman}
\affiliation{Department of Astronomy and Astrophysics,
The Pennsylvania State University, University Park, PA 16802, USA;
kll207@psu.edu}
\affiliation{Center for Exoplanets and Habitable Worlds, The
Pennsylvania State University, University Park, PA 16802, USA}

\author{C. Alves de Oliveira}
\affiliation{European Space Agency, European Space Astronomy Centre,
Camino Bajo del Castillo s/n, 28692 Villanueva de la Ca\~{n}ada, Madrid, Spain}

\author{I. Baraffe}
\affiliation{Physics \& Astronomy Department, University of Exeter, Exeter,
EX4 4QL, UK}
\affiliation{Ecole Normale Sup\'{e}rieure de Lyon, CRAL, CNRS UMR 5574, 
69364, Lyon Cedex 07, France}

\author{G. Chabrier}
\affiliation{Physics \& Astronomy Department, University of Exeter, Exeter,
EX4 4QL, UK}
\affiliation{Ecole Normale Sup\'{e}rieure de Lyon, CRAL, CNRS UMR 5574, 
69364, Lyon Cedex 07, France}

\author{E. Manjavacas}
\affiliation{AURA for the European Space Agency, Space Telescope Science
Institute, 3700 San Martin Drive, Baltimore, MD 21218, USA}
\affiliation{Department of Physics \& Astronomy, Johns Hopkins University,
Baltimore, MD 21218, USA}

\author{R. J. Parker}
\affiliation{Department of Physics and Astronomy, The University of Sheffield, 
Hicks Building, Hounsfield Road, Sheffield S3 7RH, UK}
\affiliation{Royal Society Dorothy Hodgkin Fellow}

\author{P. Tremblin}
\affiliation{Universit\'{e} Paris-Saclay, UVSQ, CNRS, CEA,
Maison de la Simulation, 91191, Gif-sur-Yvette, France}

\begin{abstract}

We have used the multiobject mode of the Near-Infrared Spectrograph 
(NIRSpec) on board the James Webb Space Telescope (JWST) to obtain
low-resolution 1--5~\micron\ spectra of 22 brown dwarf candidates
in the Orion Nebula Cluster, which were selected with archival images
from the Hubble Space Telescope. One of the targets was previously classified
as a Herbig-Haro (HH) object and exhibits strong emission in H~I, H$_2$,
and the fundamental band of CO, further demonstrating that HH objects can have
bright emission in that CO band. The remaining targets have late spectral types
(M6.5 to early L) and are young based on gravity sensitive features,
as expected for low-mass members of the cluster. 
According to theoretical evolutionary models, these objects should have masses 
that range from the hydrogen burning limit to 0.003--0.007~$M_\odot$.
Two of the NIRSpec targets were identified as proplyds in earlier analysis
of Hubble images. They have spectral types of M6.5 and M7.5,
making them two of the coolest and least massive known proplyds.
Another brown dwarf shows absorption bands at 3--5~\micron\ from
ices containing H$_2$O, CO$_2$, OCN$^-$, and CO, indicating that it
is either an edge-on class II system or a class I protostar.
It is the coolest and least massive object that has detections of these
ice features. In addition, it appears to be the first candidate 
for a protostellar brown dwarf that has spectroscopy confirming its late
spectral type.

\end{abstract}

\section{Introduction}
\label{sec:intro}

The Orion Nebula Cluster \citep[ONC,][]{mue08,ode08} is one of the most 
attractive sites for studying the formation of brown dwarfs.
Because of the youth \citep[1--5 Myr,][]{jef11} and proximity 
\citep[$390\pm2$ pc,][]{mai22} of the ONC, its substellar members can be 
detected down to very low masses.
The richness of the cluster ($\sim$2000 members) allows good statistical
constraints on the mass function and other properties of its brown dwarfs
(e.g., binary fraction, disk fraction).
The high stellar density and resulting compactness on the sky of the ONC
make it amenable to deep imaging and multiobject spectroscopy to identify
and characterize the substellar members.
In addition, the ONC can be compared to sparser star-forming regions
to search for variations with stellar density in the mass function of
brown dwarfs, and perhaps the occurrence rate of ejected giant planets
\citep{smi01,hur02,fla19,daf22}.
The primary challenge in observing brown dwarfs in the ONC is the bright
emission from the Orion Nebula, which reduces the sensitivity of images
and spectroscopy. 
UV radiation from the most massive O star in the ONC, 
$\theta^1$ C Ori, is ultimately responsible for most of that emission
through the generation of an H~II region and the heating of the surrounding
molecular cloud. The UV photons and the H~II region do offer a distinct 
advantage in studying the members of the ONC, making it possible for 
high-resolution imaging to spatially resolve ionization fronts surrounding 
circumstellar disks (known as ``proplyds") and the silhouettes of disks 
against the background nebular emission.
Many examples of these disks were discovered by the Hubble Space Telescope 
(HST) in its early years of operation 
\citep{ode93,ode94,ode96,mcc96,bal00,ric08}.

Surveys for brown dwarfs in the ONC have detected 
several hundred candidates
using optical and near-infrared imaging with ground-based telescopes
\citep{mcc94,sim99,luc00,hil00,mue02,luc05,rob10,dar12,dra16,mei16}
and HST \citep{luh00,and11,rob20,gen20}.
A small fraction of the candidates have been observed with ground-based 
spectroscopy to confirm their youth and late spectral types
\citep{hil97,luc01,luc06,sle04,rid07,wei09,hil13,ing14}.
Given its unmatched infrared (IR) sensitivity and wide array of camera filters, 
the James Webb Space Telescope \citep[JWST,][]{gar23} can identify brown 
dwarfs in the ONC more reliably and at lower masses than previous facilities.
In particular, its high angular resolution greatly facilitates the detection
of faint sources within the bright, spatially variable emission from the
Orion Nebula \citep{mcc23}. In addition, the Near-Infrared Spectrograph 
\citep[NIRSpec,][]{jak22} on JWST has a multiobject mode that can
observe $>10$ ONC members simultaneously with the optimal sensitivity
of slit spectroscopy \citep{fer22}. 
To demonstrate this capability, we have performed NIRSpec
observations of a sample of brown dwarf candidates in the ONC that were
selected from HST images \citep{rob20}. This paper presents those data,
which include proplyds near and below the hydrogen burning
limit and a candidate for a protostellar brown dwarf.

\section{Spectroscopy}

\subsection{NIRSpec Observations}

We pursued spectroscopy of brown dwarf candidates in the ONC with NIRSpec on 
JWST through guaranteed time observation program 1228 (PI: C. Alves 
de Oliveira). The observations utilized NIRSpec's microshutter assembly (MSA),
which consists of four quadrants that each contain 365$\times$171 shutters. 
The angular size of an individual shutter is $0\farcs2\times0\farcs46$.
The MSA spans a field with a size of $3\farcm6\times3\farcm4$.
The data are collected by two $2048\times2048$ detector arrays that have
pixel sizes of $0\farcs103\times0\farcs105$. We selected the PRISM disperser 
for the ONC observations because it provides the best sensitivity and the 
widest wavelength coverage (0.6--5.3~\micron) among the available options. 
The spectral resolution for PRISM data ranges from $\sim$40 to 300 from
shorter to longer wavelengths. 

The targets for NIRSpec were selected from near-IR images obtained 
with the Wide Field Camera 3 on HST \citep[WFC3,][]{kimb08} in 2015
through Treasury program 13826 (PI: M. Robberto). These images were 
taken in a filter that coincides with absorption bands of H$_2$O and CH$_4$ 
that are strong in brown dwarfs \citep{kir05}
and a filter that measures the neighboring continuum, corresponding
to F139M and F130N, respectively.
Measurements of astrometry and photometry for sources detected in the
WFC3 images are available from \citet{rob20}, but we utilized a
reduction of the data that we performed in 2017 during the initial 
preparations for the NIRSpec observations. We considered only the WFC3 images 
that are within a radius of $\sim4\arcmin$ from the center of the ONC.

\begin{figure}
\epsscale{1.2}
\plotone{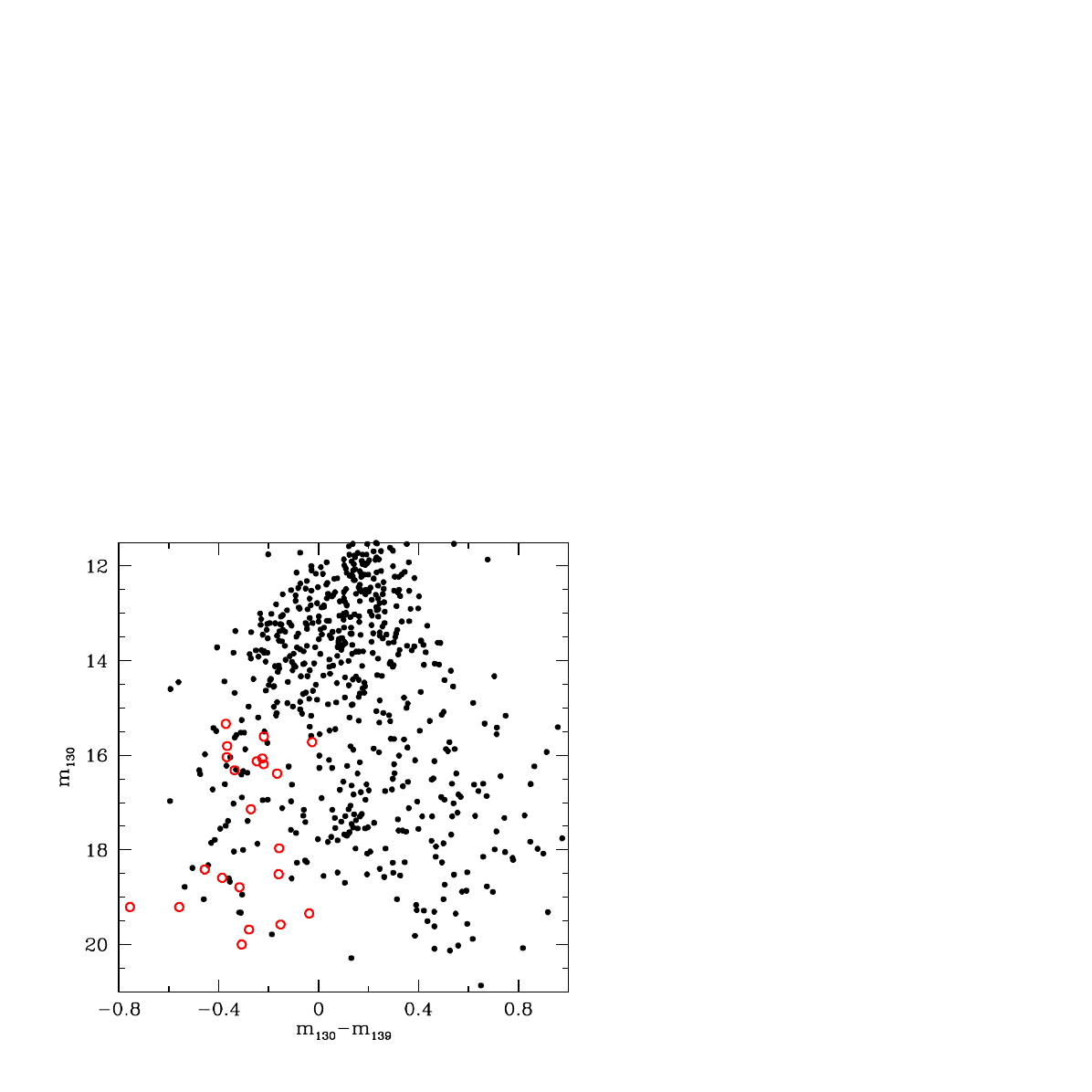}
\caption{
Color-magnitude diagram for sources within the field shown in
Figure~\ref{fig:map} based images from HST/WFC3 \citep{rob20}.
We have obtained JWST/NIRSpec data for a sample of brown dwarf candidates
that have $m_{130}-m_{139}<0$ and $m_{139}>15$ (circles).}
\label{fig:cmd}
\end{figure}

\begin{figure}
\epsscale{1.2}
\includegraphics[width=\textwidth]{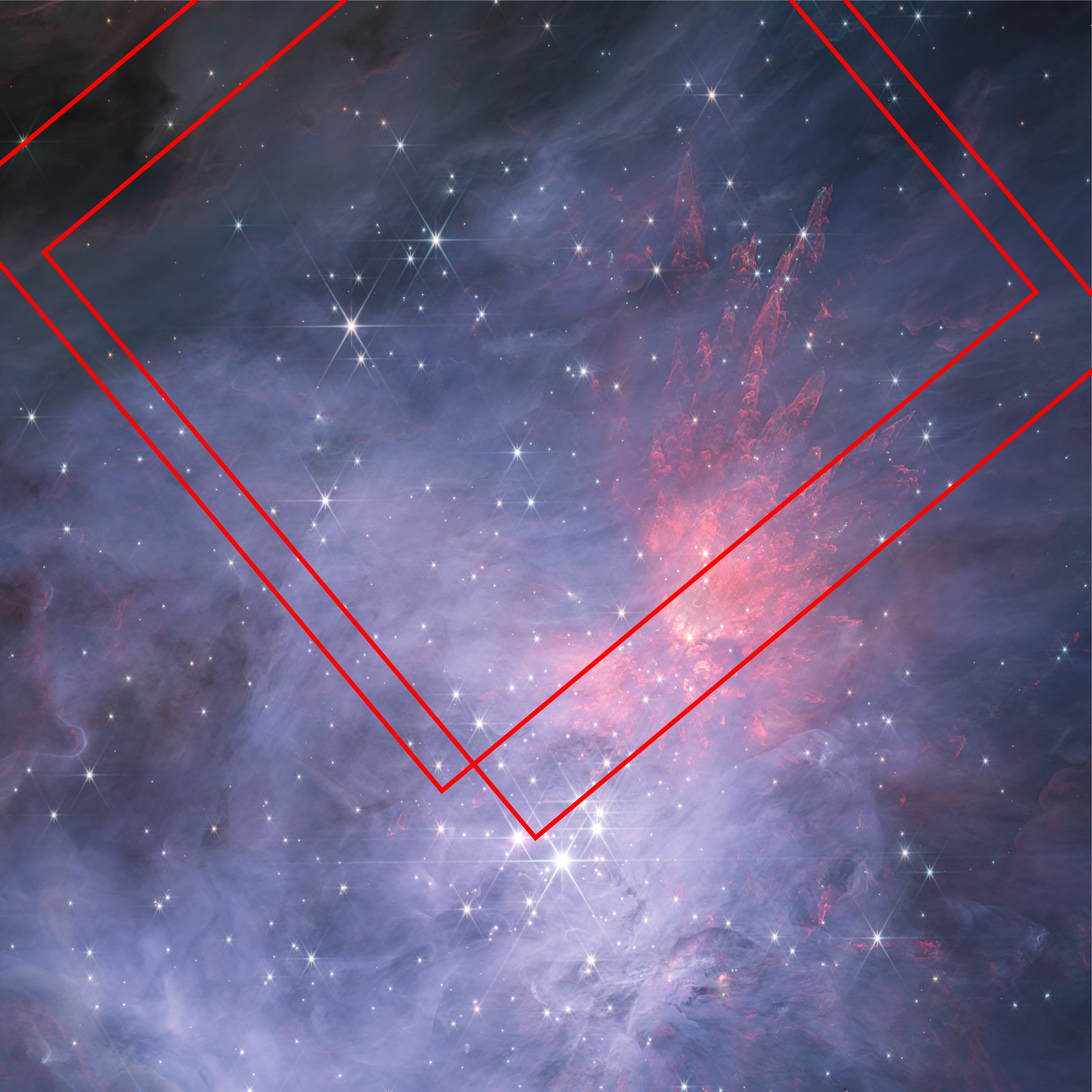}
\caption{
JWST/NIRCam images in five filters (F140M, F162M, F182M, F187N, 
F212N) for a $5\arcmin\times5\arcmin$ field in the center of the 
ONC \citep{mcc23}. We have indicated the field of 
view for the NIRSpec MSA at two pointings (rectangles).
The fluxes are displayed on a logarithmic scale.
North is up and east is left.}
\label{fig:map}
\end{figure}

In Figure~\ref{fig:cmd}, we have plotted a diagram of $m_{130}$ versus
$m_{130}-m_{139}$ for the WFC3 sources that are within the
$5\arcmin\times5\arcmin$ field near the center of the ONC that is shown
in Figure~\ref{fig:map}. At progressively fainter magnitudes, members of the
ONC are expected to have lower temperatures and increasing absorption in the 
bands overlapping with F139M, leading to bluer values of $m_{130}-m_{139}$
\citep{rob20}. As a result, the bluest sources at fainter magnitudes
are the most promising candidates for substellar members of the ONC.
Faint sources with redder colors should be a mixture of highly reddened
brown dwarfs, stellar members that are seen in scattered light (edge-on
disks and protostars), and background objects. 
The number of sources quickly increases with redder
colors beyond $m_{130}-m_{139}\sim0$, which likely reflects contamination
from background stars and galaxies \citep{rob20}, so we selected sources that
have $m_{130}-m_{139}<0$ for our sample of potential NIRSpec targets. 
We also required $m_{130}>15$ to avoid saturation in NIRSpec. 
Among those with $m_{130}-m_{139}<-0.15$, higher priorities were
assigned at fainter magnitudes. The candidates at $-0.15<m_{130}-m_{139}<0$
served as filler targets, which had the lowest priority.

We used the MSA Planning Tool (MPT) in the Astronomer's Proposal Tool (APT)
to design two MSA configurations for observing a subset of the brown dwarf 
candidates identified with WFC3.
We selected settings in MPT such that each target would be observed in three 
adjacent shutters that are equivalent to a single $0\farcs2\times1\farcs5$ 
slitlet, and each target would have a separation of $\geq0\farcs059$ 
from the edges of the shutter. 
We required that the two MSA configurations would have a separation
less than the visit splitting distance (65$\arcsec$) so that they could be
observed in single visit. We used MPT to search for a pair of configurations
that would (1) maximize the number of targets, particularly those at higher
priorities, (2) avoid overlap of the MSA with the Trapezium stars since
the shutters are not completely opaque, and (3) have suitable target 
acquisition stars (no close neighbors, not located in the brightest nebular
emission, appropriate magnitudes). The targets for the selected MSA 
configurations are located north of the Trapezium stars.  
In Figure~\ref{fig:map}, we show 1.4--2.1~\micron\ images for a field 
surrounding the MSA configurations obtained with NIRCam on JWST 
\citep{rie05,rie23} by \citet{mcc23}. We have marked the field of view of the 
MSA for each pointing. Although we avoided the 
Trapezium stars, one of MSA configurations did overlap with the 
Becklin-Neugebauer (BN) object \citep{bec67}, which is faint in WFC3 bands but 
is very red and becomes extremely bright longward of 2~\micron.  It is the 
bright source at the center of the red nebula in Figure~\ref{fig:map}.
Enough light from BN leaked through the closed shutters of the MSA to
contaminate the data for two targets (Section~\ref{sec:reduction}).

Fourteen targets were observed in each of the two MSA configurations.
Three sources were present in both configurations, so the total number of 
targets was 25.
The NIRSpec data for three targets were not useful
(Section~\ref{sec:reduction}). 
The remaining 22 sources are listed in Table~\ref{tab:spec}, which includes 
the source numbers from the MPT catalog, equatorial coordinates
measured from the NIRCam images, spectral types, and extinction estimates.
One of the targets, source 143 ([OW94] 130$-$119), 
was previously classified as a Herbig-Haro (HH) object \citep{lee00}, and thus 
should have been omitted from our sample of brown dwarf candidates.
Four of the targets had filler status during the design of the MSA observations.
NIRSpec reveals one to be a proplyd near the hydrogen burning 
limit (source 69, Section~\ref{sec:pro}) and the other to be a brown dwarf that
may be protostellar (source 126, Section~\ref{sec:ice}).

The observations with the two MSA configurations were performed on
2023 February 22 (UT).  At each of the three nod positions, one exposure
was taken, which utilized five groups, six integrations, and the NRSRAPID
readout pattern. The latter was chosen to avoid saturation for the
brightest targets. In a given MSA configuration, the total exposure time 
for three exposures was 1160~s. The charged time was 2.58~hrs.

\subsection{Data Reduction}
\label{sec:reduction}

To reduce the NIRSpec data, we began by retrieving the {\tt uncal}
files from the Mikulksi Archive for Space Telescopes (MAST):
\dataset[doi:10.17909/zj8z-bs59]{http://dx.doi.org/10.17909/zj8z-bs59}.
Those files were processed in the manner described by \citet{luh24}
for NIRSpec observations in IC 348. That study utilized 
a custom version of the pipeline developed by the ESA NIRSpec Science
Operations Team \citep{fer22} and based on the workflow and algorithms 
described in \citet{alv18}.

\begin{figure}
\epsscale{1.2}
\plotone{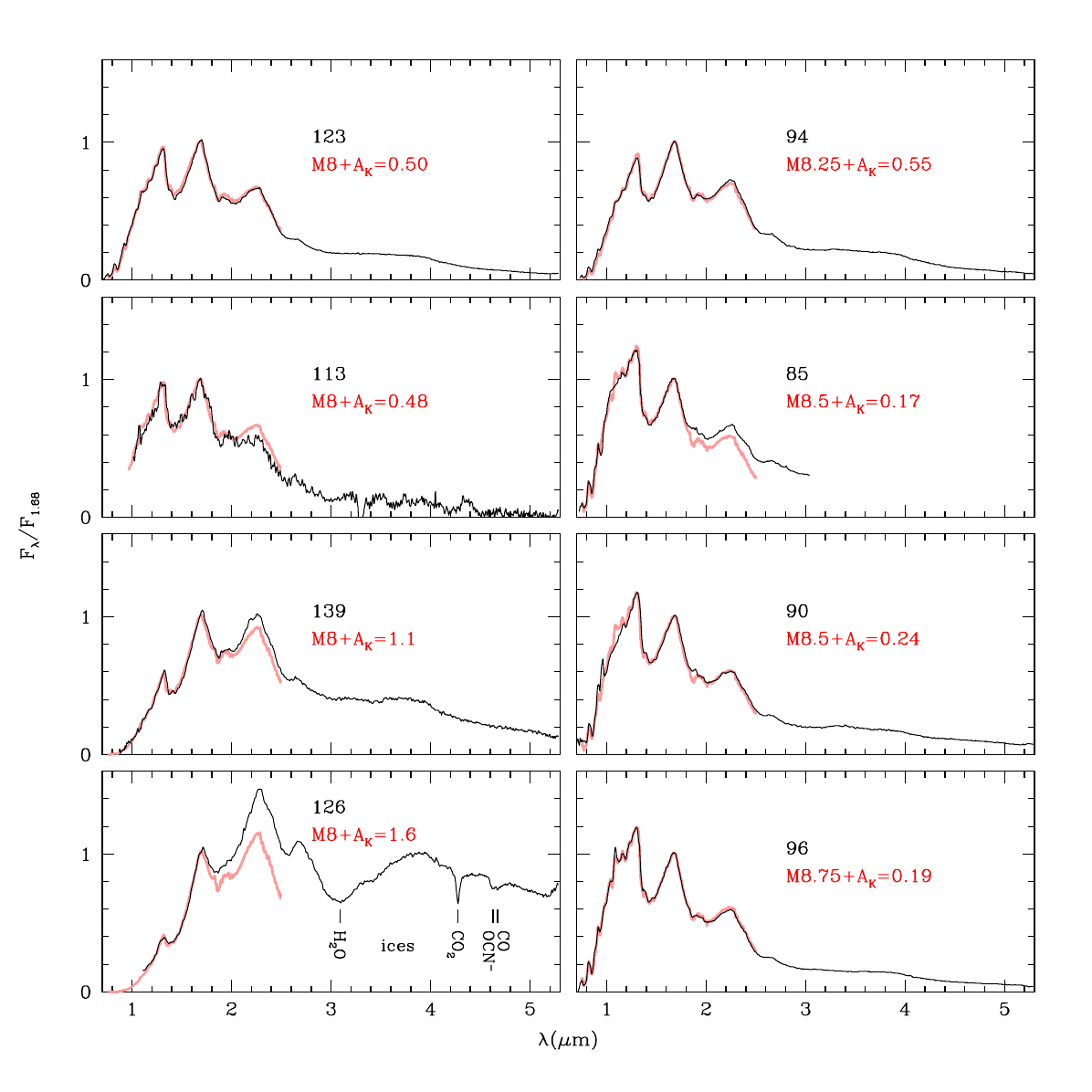}
\caption{
More JWST/NIRSpec spectra in the ONC.}
\label{fig:spec2}
\end{figure}

\begin{figure}
\epsscale{1.2}
\plotone{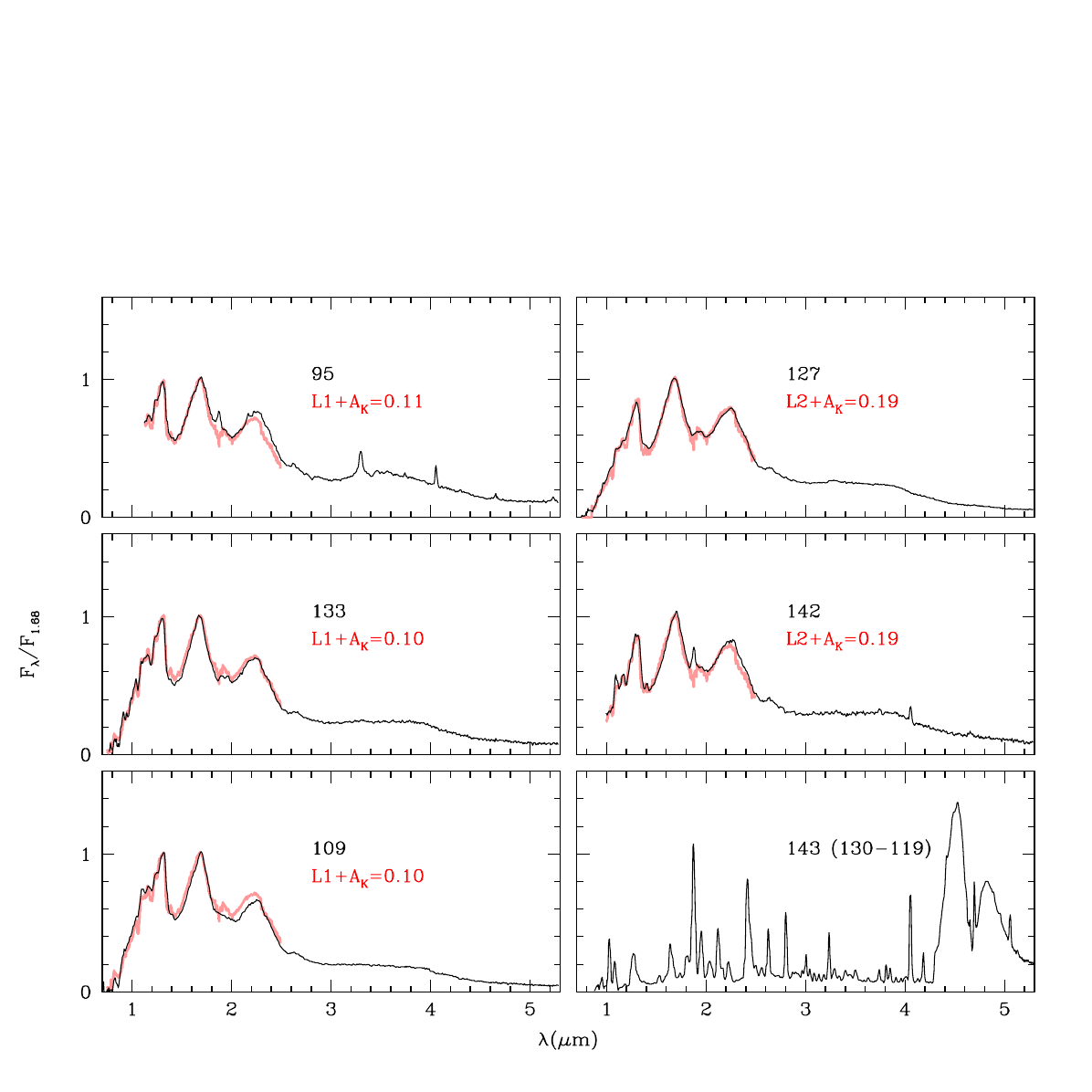}
\caption{
More JWST/NIRSpec spectra in the ONC.}
\label{fig:spec3}
\end{figure}

Since the NIRSpec data were collected at three nod positions in each
MSA configuration, the background subtraction for a given nod can utilize
the average of the other two nods or only one of them.
The average was our default choice, but for a few nods, we omitted one 
of the other nods in the background subtraction because of the presence of
a second source or artifact in the slit or a variation in the background 
emission along the slit. For five sources (80, 85, 95, 113, 143), one of 
the nods was not used because neither of the other two nods produced adequate
background subtraction. After subtraction, some of the nods
exhibited small positive or negative residuals at the wavelengths
of nebular emission lines, which are likely caused by
spatial variations in the background emission. 
We interpolated the continua of the spectra across
these residuals. The weaker emission lines that remain in the spectra
may arise from either the sources or the nebula.
For a few objects, all of the nods contained strong emission 
lines with similar strengths, indicating that the lines are concentrated
on those objects. We have ignored the data for source 85 longward of 
3~\micron\ because of contamination by light from BN leaking through the MSA.
For three sources, the wavelength ranges falling on the detectors 
were too limited to be useful, so we do not report their spectra.
One of those objects also was affected by artifacts related to BN.
The reduced NIRSpec data for the remaining 22 targets are presented in 
Figures~\ref{fig:spec1}--\ref{fig:spec3}.
The signal-to-noise ratios are $>50$ for most wavelengths and objects.
The spectrum of source 113 has a negative residual near 3.3~\micron\ that 
corresponds to the brightest emission feature from polycyclic aromatic 
hydrocarbons and reflects imperfect background subtraction.

\subsection{Spectral Classifications}
\label{sec:spt}

Source 143 ([OW94] 130$-$119) was identified as an HH object based on
narrowband imaging in transitions of H$_2$ and [Fe II] \citep{lee00}. 
Its NIRSpec spectrum is dominated by emission in H~I, H$_2$, and the 
fundamental band of CO. Previous studies of broadband mid-IR images have 
suggested that some HH objects produce bright CO emission \citep{tak10,tap12}, 
which was recently confirmed with NIRSpec \citep{ray23}. 
Our data for source 143 provide additional evidence of that kind.

The remaining NIRSpec targets exhibit (1) strong H$_2$O absorption
bands that indicate spectral types of late M through L and (2) gravity
sensitive features that indicate young ages, which 
include the triangular shape of H-band continuum \citep{luc01} and 
the weak CO band at 4.4--5.2~\micron\ \citep{luh23}. 
The youth of the targets supports their membership in the ONC.

We have estimated spectral types and reddenings through 
visual comparison of the NIRSpec data at $<2.5$~\micron\ to standard
spectra for young stars and brown dwarfs \citep[M0--L7,][]{luh17}. 
For each spectral type in the range of the standards, 
progressively larger reddening \citep{sch16av} was applied to the standard until
it matched the slope of a given NIRSpec target. From among those
reddened standards, we then identified the spectral type that provided
the best match to the absorption features.
The resulting types range from M6.5 to early L.
In Figures~\ref{fig:spec1}--\ref{fig:spec3}, we show the NIRSpec data in
order of spectral type. Each spectrum is compared to a standard for
the best matching spectral type and reddening.  
The standards are reddened to match the ONC objects at
1.2--1.7~\micron, which usually produces good matches to the entire slopes at
$<2.5$~\micron. A few objects (69, 87, 126) 
have large excesses in the $K$-band relative to the standards.
They also show excesses at longer wavelengths relative to the bluest 
(photospheric) sources at similar spectral types (Section~\ref{sec:exc}), 
indicating the presence 
of increasing emission at longer wavelengths from circumstellar dust.
For the sources with $K$-band excesses, non-negligible disk emission could
extend to shorter wavelengths and weaken the H$_2$O band at 1.4~\micron,
resulting in spectral type estimates that are slightly too early.
The coolest objects are consistent with wider ranges of spectral types 
(e.g., reddened late-M to unreddened L), reflecting degeneracies between 
spectral type and reddening in low-resolution near-IR spectra of 
young L dwarfs \citep{luh17}.

Source 113 is one of our faintest targets, but it is not among the coolest
according to our classification (M8). As a result, it is unusually faint
for its spectral type among members of the ONC. Similarly, it appears
underluminous in color-magnitude diagrams constructed from NIRCam
photometry for the cluster. This characteristic of source 113 may
indicate that it is observed primarily in scattered light.
Many underluminous sources have been previously found during
spectroscopy of brown dwarf candidates in the ONC \citep{sle04}.

Previous studies have reported spectral types for two of our targets,
sources 96 \citep[M7.5$\pm$1.5,][]{sle04} and 120 \citep[M3.75--M5,][]{ing14}.
Our classifications are M8.75 and M6.5, respectively. Source 120 is
$1\farcs7$ from a brighter and earlier star, which may have contaminated
the previous ground-based spectrum.

It is useful to consider the ONC sources in the context of the
coolest young brown dwarfs that have been observed spectroscopically 
with NIRSpec. None of our targets are as cool as the L-type companion TWA 27B 
\citep{cha04} based on comparisons to its NIRSpec data \citep{luh23}.
NIRSpec has recently discovered absorption features from an unidentified 
aliphatic hydrocarbon in two members of IC 348 \citep{luh24}.
These features are not detected in any of the ONC targets, which is not
surprising since their photometry would suggest that they probably do
not extend down to the masses and temperatures of the objects in IC 348.

\section{Properties of Brown Dwarfs}

\subsection{Mass Estimates}
\label{sec:mass}

The earliest spectral type among the NIRSpec targets is M6.5, which
should correspond to a mass near the hydrogen burning limit based
on temperature scales for young stars \citep{luh03} and
theoretical evolutionary models \citep{bar98,bar15,cha23}.
The extinction-corrected magnitudes of the targets are as faint as
$m_{130}\sim19.6$. For that limit, the models 
from \citet{cha23} suggest masses of 0.003--0.007~$M_\odot$ assuming 
an age range of 1--5 Myr for members of the ONC \citep{jef11}.

\subsection{Proplyds}
\label{sec:pro}

Two of the NIRSpec targets, source 69 ([OW94] 181$-$247) and source 117 
([OW94] 131$-$046), have been previously identified as proplyds
based on optical images from HST \citep{ode96}. Those data included 
filters that measure emission lines of H~I, N~II, and O~III, which
trace the ionization front in a proplyd.  Each object exhibits a tear-drop
shaped nebula with a tail that points away from $\theta^1$ C Ori.
Dust continuum emission at millimeter wavelengths combined
with a standard gas-to-dust ratio implies disk masses of $6.2\pm1.0$ and 
$2.2\pm0.4$ $M_{\rm Jup}$ for sources 69 and 117, respectively \citep{eis16}.

\begin{figure}
\epsscale{1.2}
\plotone{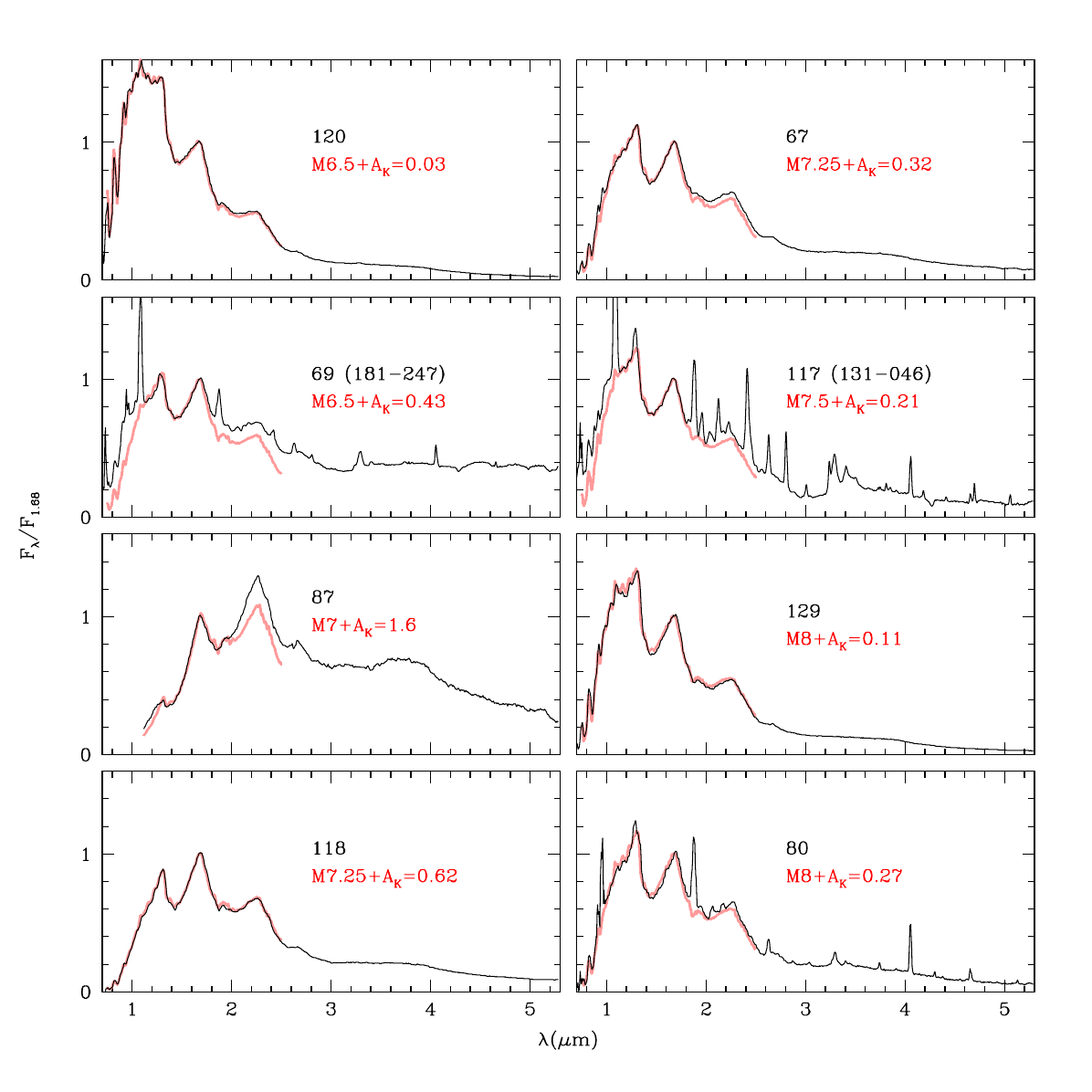}
\caption{
JWST/NIRSpec spectra of brown dwarf candidates in the ONC.
The spectra are labeled with the source numbers from Table~\ref{tab:spec}
and are compared to young standard spectra that have been reddened
to match the slopes from 1.2--1.7~\micron\ \citep{luh17}.}
\label{fig:spec1}
\end{figure}

As expected, the NIRSpec data for the two proplyds contain strong emission 
lines, most notably from H~I and He~I (Figure~\ref{fig:spec1}).
In addition, the spectra exhibit absorption bands from H$_2$O that
indicate late spectral types. In Section~\ref{sec:spt}, we estimated
types of M6.5 and M7.5 for sources 69 and 117, respectively.
It is difficult to estimate masses of central stars in proplyds from photometry
because of reddening, line emission, and scattered light, but these
spectral types should correspond to masses near and below the hydrogen burning 
limit (Section~\ref{sec:mass}).
Source 69 appears to have $K$-band excess emission relative to young
standard spectra (Figure~\ref{fig:spec1}), so its true spectral type
could be slightly later than our estimate (Section~\ref{sec:spt}).

\begin{figure}
\epsscale{1.2}
\plotone{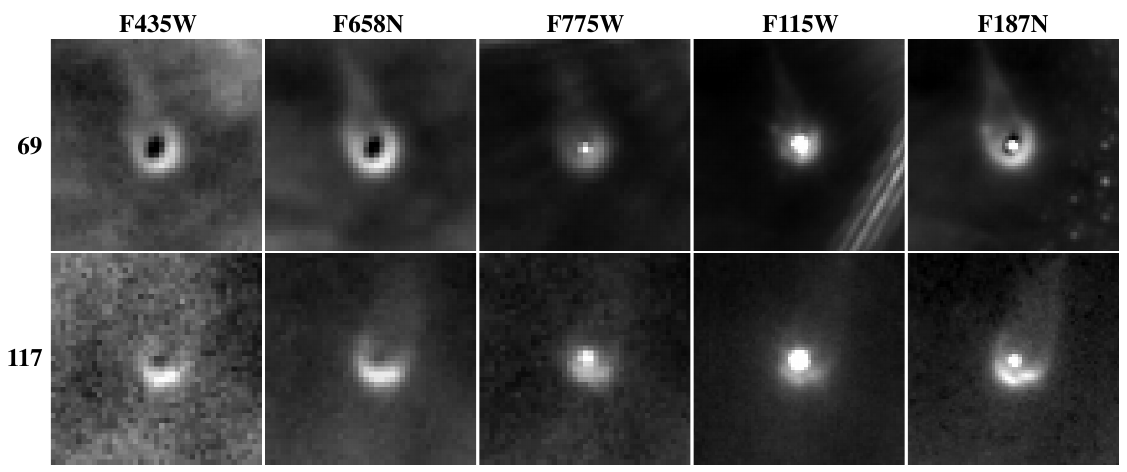}
\caption{
HST/ACS and JWST/NIRCam images of two ONC proplyds observed by NIRSpec
\citep{ric08,rob13,mcc23}. The size of each image is $2\arcsec\times2\arcsec$. 
North is up and east is left.}
\label{fig:prop}
\end{figure}

Since the discovery of sources 69 and 117, additional high-resolution images
have been taken with the Advanced Camera for Surveys (ACS) on HST 
\citep{ric08,rob13} and NIRCam on JWST \citep{mcc23}. In Figure~\ref{fig:prop}, 
we show images for a selection of the available filters that illustrate the 
colors of the central sources and that detect the ionization fronts,
consisting of three broadband filters (F435W, F775W, F115W) and
two narrowband filters that measure H$\alpha$+[N~II] (F658N) and 
Paschen $\alpha$ (F187N).  Each of the central sources is absent in F435W 
and F658N and is detected in the remaining bands, becoming brighter at longer 
wavelength, which reflects a combination of late spectral types and
extinction (Figure~\ref{fig:spec1}).
For source 69, the area within the ionization front is darker in F435W and
F658N than the background emission that surrounds the proplyd, which we
interpret as a silhouette of the disk. As the flux of the central
source increases at longer wavelengths, its point spread function makes
it more difficult to detect a silhouette, but one may be present in F187N
as well. The radius of the silhouette is $\sim0\farcs08$, corresponding to
$\sim$30 AU at the distance of the ONC. Source 117 may also contain a smaller
disk silhouette in F435W and F658N.

One additional NIRSpec target, source 80, has strong emission
lines that are suggestive of a proplyd (Figures~\ref{fig:spec1}.
It is a point source in the available images from HST and JWST, so it
may have an ionization front that is too small to resolve.

Two candidates for substellar proplyds have been highlighted in
previous work, both of which are located in the ONC.
One of the objects is [OW94] 124$-$132, which consists of a $0\farcs15$ 
binary system that has a silhouette disk at shorter wavelengths
and is surrounded by an ionization front \citep{ode96,smi05,rob08}.
\citet{rob08} proposed that the components of the binary are substellar
based on their photometry. However, spectroscopy of the system
is not available, and its relatively red color of $m_{130}-m_{139}=0.14$ 
\citep{rob20} appears to be inconsistent with a late spectral type 
for its estimated extinction \citep{rob08}.
The second previous candidate for a brown dwarf proplyd 
is [OW94] 133$-$353 \citep{ode96,ric08,fan16}, which is one component of a 
$1\arcsec$ pair. In HST and JWST images, it appears as a point source
with extended emission pointed away from $\theta^1$ C Ori.
The direction and cometary shape of the extended emission are suggestive
of a proplyd. If an ionization front surrounds the central source, it is
unresolved. \citet{fan16} performed spectroscopy on the object, measuring
a spectral type of M9.5. Its $m_{130}-m_{139}$ color is also sufficiently
blue to indicate a late spectral type \citep{rob20}.
With a magnitude of $m_{130}=14.4$, it is unusually bright for its
spectral type among members of the ONC.

Dozens of additional proplyds in the ONC are faint enough to be substellar
\citep{ric08}, but they also could be stars that are observed through 
high extinction or primarily in scattered light. Only a few of these objects 
have spectral classifications, which include
[OW94] 179$-$354 and [OW94] 044$-$527 \citep{bal98,bal00}. 
The former has optical and IR types of K8--M3 and M7.5:, respectively
\citep{sle04}. New spectroscopy would be useful to resolve these discrepant
types. The second source has been classified as M8.25--M8.5 \citep{wei09,ing14},
which should place it well below the hydrogen burning limit.

\subsection{Excess Emission from Disks}
\label{sec:exc}

The NIRSpec data extend to long enough wavelengths to potentially
encompass emission from dust in circumstellar disks. 
Therefore, we can use these data to search for evidence of disks.
The cool dust in a disk generates IR continuum emission that appears as an 
excess above that expected from the stellar photosphere. The size of the
excess grows with longer wavelengths, causing the observed spectrum of the
system to be redder than the photosphere alone.
Our sample is large enough that it should contain both diskless and 
disk-bearing objects, so we have compared the spectral slopes among
the targets to identify the two populations.
For this analysis, the spectra are corrected for the extinctions
derived during the spectral classifications and are normalized
at 1.68~\micron, as done for the observed spectra in 
Figures~\ref{fig:spec1}--\ref{fig:spec3}. In addition, the comparisons
are performed for groups of objects with similar spectral types 
since photospheric spectral slopes vary with type.
Near a given spectral type, most of the spectra have similar slopes,
spanning a range of $\sim0.1$~mag at 3--5~\micron\ when normalized
at 1.68~\micron, and a minority are significantly redder, which correspond to 
stellar photospheres and disk-bearing sources, respectively.
In Table~\ref{tab:spec}, we indicate whether we find excesses for each
of the targets. Three objects have large IR excesses 
($>$0.5 mag at 3--5~\micron),
consisting of sources 69, 87, and 126. A few additional
sources may have small excesses ($\sim0.2$ mag), but data at 
longer wavelengths are needed for confirming the presence of disks.
Their IR excess flags are listed as ``?" in Table~\ref{tab:spec}.
As mentioned in Section~\ref{sec:spt}, the cooler objects in our sample
are consistent with wider ranges of types, so their extinctions
and photospheric colors are uncertain, making it more difficult
to reliably identify excesses from disks. 
We do not find large excesses for any of those objects,
but smaller excesses are possible. Therefore, their IR excess 
flags appear as ``no?" in Table~\ref{tab:spec}.
Disks around brown dwarfs often do not produce significant excesses at
wavelengths shortward of 5~\micron\ \citep{luh10tau}, so the objects that
lack large excesses in the NIRSpec data can still have disks, as in the 
case of one of the proplyds in our sample, source 117.

\subsection{Candidate for Edge-on Disk or Protostar}
\label{sec:ice}

One of the NIRSpec targets with IR excess emission, source 126, is distinctive 
in that it shows absorption bands near 3 and 4.27~\micron\ and a blend of two 
bands centered near 4.62 and 4.67~\micron\ (Figure~\ref{fig:spec2}).
These features have been observed in edge-on disks and class I protostars
and have been attributed to ices that contain H$_2$O, CO$_2$, OCN$^-$, 
and CO, respectively \citep{thi02,aik12,boo22,kim22,stu23,fed24}.
Given its high extinction ($A_K=1.6$), the object could be either an edge-on
class II system that is reddened by the molecular cloud or a
class I protostar\footnote{The stages of a young stellar object consist
of classes 0 and I (protostar+disk+infalling envelope), class II (star+disk),
and class III (star) \citep{lw84,lad87,and93,gre94}.}.
Measurements of the spectral energy distribution at IR wavelengths longward
of the NIRSpec data are needed distinguish between these two possibilities,
but they are not available. To our knowledge, source 126 is the first object
in any star-forming region that has been spectroscopically 
classified to be cool enough to be substellar and that has detections of 
these ice features.

One edge-on disk has been previously detected around a brown dwarf
\citep{luh07}, which has silicate absorption at 10~\micron\ from the
occulting disk. It likely has the same ice features observed in source 126,
but it has not been observed spectroscopically at those wavelengths.
Meanwhile, a few dozen candidates for class 0 and I brown dwarfs have been
identified \citep{har99,ria17}, primarily with the Spitzer Space Telescope 
\citep{you04,bou06,dun08,bar09}. They appear to have luminosities low 
enough to be substellar based on their IR spectral energy distributions, 
but most lack the spectroscopy needed for confirmation of late spectral types. 
For the protostellar candidates that do have spectra, the spectral
classifications have been indicative of stellar masses \citep{luh10,luh22}.
Source 126 appears to be the first candidate for a protostellar brown dwarf that
has been shown to have a late spectral type.

\section{Conclusions}

We have used JWST/NIRSpec to perform multi-object spectroscopy
of brown dwarf candidates in the ONC, which were selected from
HST/WFC3 images \citep{rob20}. Our results are summarized as follows:

\begin{enumerate}

\item
We have observed 22 brown dwarf candidates near the center of the ONC 
with two MSA configurations of NIRSpec. 
Data were collected with the PRISM disperser, which provided a
wavelength coverage of 0.6--5.3~\micron\ and a resolution ranging 
from $\sim$40 to 300.

\item
One of the targets, [OW94] 130$-$119, was previously classified
as a HH object \citep{lee00}. Its NIRSpec spectrum exhibits 
strong emission in H~I, H$_2$, and the fundamental band of CO.
As with other recent NIRSpec observations \citep{ray23}, our data
confirm that HH objects can produce bright emission in that CO band.

\item
The remaining NIRSpec targets are cool and young based on 
strong H$_2$O absorption bands and gravity sensitive features, as expected
for substellar members of the ONC. We have estimated spectral types
and extinctions from these data through comparison to young standard spectra.
The resulting types range from M6.5 to early L. Based on theoretical
evolutionary models \citep{cha23}, the earliest object should have
a mass near hydrogen burning limit while the faintest and coolest objects
have mass estimates of 0.003--0.007~$M_\odot$ for ages of 1--5 Myr.

\item
Two of the NIRSpec targets are known proplyds based on images from HST 
\citep{ode96}. We have classified them as M6.5 and M7.5 with NIRSpec,
making them two of the coolest and least massive known proplyds.
In optical images from HST \citep{ric08}, the M6.5
proplyd appears to exhibit a silhouette disk that has a radius of 
$\sim0\farcs08$ ($\sim$30 AU).

\item
For three objects in our sample, the NIRSpec data are
significantly redder than expected for reddened photospheres, indicating
the presence of IR excess emission from disks. 
One of those sources also shows absorption bands
at 3--5~\micron\ from ices containing H$_2$O, CO$_2$, OCN$^-$, and CO,
which have been previously observed in edge-on disks and protostars.
Given its high extinction, it could be either an edge-on
class II system that is reddened by the molecular cloud or a
class I protostar. To our knowledge, it is the first object
spectroscopically classified as a brown dwarf that has detections of 
these ice features. In addition, it appears to be the first candidate 
for a protostellar brown dwarf that has spectroscopy confirming its late
spectral type.

\item
Our study has demonstrated the ability of JWST/NIRSpec to perform multiobject
spectroscopy on brown dwarf candidates within the bright extended emission 
of the Orion Nebula.
Given that capability and the high yield of confirmed brown dwarfs in
our sample, it should be feasible to obtain spectra of a large,
representative sample of brown dwarfs in the ONC that extends down to
masses of $\sim1$~$M_{\rm Jup}$.

\end{enumerate}

\begin{acknowledgments}

P.T. acknowledges support from the European Research Council under grant 
agreement ATMO 757858.  R.P. acknowledges support from the Royal Society in 
the form of a Dorothy Hodgkin Fellowship. The JWST data were obtained from 
MAST at the Space Telescope Science Institute, which is operated by the 
Association of Universities for Research in Astronomy, Inc., under NASA 
contract NAS 5-03127. The JWST observations are associated with program 1228.
This work made use of ESA Datalabs (\url{https://datalabs.esa.int}), which is
an initiative by ESA's Data Science and Archives Division in the Science and
Operations Department, Directorate of Science. 
The Center for Exoplanets and Habitable Worlds is supported by the
Pennsylvania State University, the Eberly College of Science, and the
Pennsylvania Space Grant Consortium.

\end{acknowledgments}

\clearpage

\begin{deluxetable}{rrlllll} 
\tabletypesize{\scriptsize}
\tablewidth{0pt}
\tablecaption{Sources in the ONC Observed with NIRSpec\label{tab:spec}}
\tablehead{
\colhead{ID\tablenotemark{a}} &
\colhead{Other Name} &
\colhead{$\alpha$ (ICRS)} &
\colhead{$\delta$ (ICRS)}  &
\colhead{Spectral Type\tablenotemark{b}}  &
\colhead{$A_K$} &
\colhead{IR Excess?}\\
\colhead{} &
\colhead{} &
\colhead{(deg)} &
\colhead{(deg)} &
\colhead{} &
\colhead{(mag)} &
\colhead{}}
\startdata
120 & \nodata &   83.783557 &   $-$5.354912 & M6.5 & 0.03 & no \\
69 & [OW94] 181$-$247 &   83.825379 &   $-$5.379750 & M6.5 & 0.43 & yes \\
87 & \nodata &   83.819875 &   $-$5.369885 & M7 & 1.6 & yes \\
118 & \nodata &   83.802672 &   $-$5.346444 & M7.25 & 0.62 & no \\
67 & \nodata &   83.831989 &   $-$5.376165 & M7.25 & 0.32 & ? \\
117 & [OW94] 131$-$046 &   83.804451 &   $-$5.346074 & M7.5 & 0.21 & ? \\
129 & \nodata &   83.814513 &   $-$5.328743 & M8 & 0.11 & no \\
80 & \nodata &   83.823234 &   $-$5.374392 & M8 & 0.27 & no \\
123 & \nodata &   83.824008 &   $-$5.324508 & M8 & 0.50 & no \\
113 & \nodata &   83.800496 &   $-$5.352120 & M8 & 0.48 & no \\
139 & \nodata &   83.847259 &   $-$5.346673 & M8 & 1.1 & no \\
126 & \nodata &   83.815390 &   $-$5.330094 & M8 & 1.6 & yes \\
94 & \nodata &   83.836852 &   $-$5.353779 & M8.25 & 0.55 & no \\
85 & IRc8 &   83.811316 &   $-$5.376508 & M8.5 & 0.17 & ? \\
90 & \nodata &   83.826230 &   $-$5.363722 & M8.5 & 0.24 & ? \\
96 & \nodata &   83.825203 &   $-$5.361406 & M8.75 & 0.19 & no \\
95 & \nodata &   83.823873 &   $-$5.362438 & M9--L2 & 0.11 & no? \\
133 & \nodata &   83.789722 &   $-$5.334200 & M9--L2 & 0.10 & no? \\
109 & \nodata &   83.814127 &   $-$5.346510 & M9--L2 & 0.10 & no? \\
127 & \nodata &   83.821114 &   $-$5.326022 & M9--L4 & 0.19 & no? \\
142 & \nodata &   83.799902 &   $-$5.325853 & M9--L4 & 0.19 & no? \\
143 & [OW94] 130$-$119 &   83.804169 &   $-$5.355232 & \nodata & \nodata & \nodata \\
\enddata
\tablenotetext{a}{Source numbers that we assigned in the MPT/APT catalog 
for JWST program 1228.}
\tablenotetext{b}{Uncertainties are 0.5 subclass unless indicated otherwise.}
\end{deluxetable}

\end{document}